\documentclass[11pt]{article}

\usepackage[top=2cm, bottom=2cm, left=2cm, right=2cm]{geometry}

\usepackage[noadjust]{cite}

\usepackage[cmex10]{amsmath}
\usepackage{amsbsy}

\usepackage{stackengine}
\usepackage{makecell}
\usepackage{boldline}
\setcellgapes{3pt}

\usepackage{algorithm}
\usepackage{algpseudocode}
\usepackage{varwidth}

\usepackage{tikz}
\usetikzlibrary{shapes, shadows, arrows}
\usepackage{url}

\usepackage{graphicx}
\usepackage{float}
\usepackage{stfloats}
\usepackage{wrapfig}

\usepackage{epstopdf}

\usepackage{amsthm,dcolumn,graphicx,multicol,fancyhdr}
\usepackage{latexsym}
\usepackage{graphicx}
\usepackage{ifthen}
\usepackage{rotating,amsfonts,color,amssymb,amsmath,amscd,array,enumerate,afterpage}

\usepackage{hyperref}
\usepackage{multirow}

\usepackage{enumitem}
\setlist{parsep=3pt,listparindent=\parindent}

\usepackage[caption=false,font=footnotesize]{subfig}

\usepackage{array}
\newcolumntype{L}[1]{>{\raggedright\let\newline\\\arraybackslash\hspace{0pt}}m{#1}}
\newcolumntype{C}[1]{>{\centering\let\newline\\\arraybackslash\hspace{0pt}}m{#1}}
\newcolumntype{R}[1]{>{\raggedleft\let\newline\\\arraybackslash\hspace{0pt}}m{#1}}

\usepackage{arydshln}

\theoremstyle{definition}

\def\subfigure{\subfloat}

\newcommand{\babc}{\renewcommand{\labelenumi}{(\alph{enumi})}\begin{enumerate}}
\newcommand{\eabc}{\end{enumerate}}
\newcommand{\biii}{\renewcommand{\labelenumi}{(\roman{enumi})}\begin{enumerate}}
\newcommand{\eiii}{\end{enumerate}}

\newcommand{\beqn}{\begin{eqnarray*}}
\newcommand{\beq}{\begin{eqnarray}}
\newcommand{\eeqn}{\end{eqnarray*}}
\newcommand{\eeq}{\end{eqnarray}}

\newcommand{\ckboldon}[1]{#1}

\newcommand{\ckbold}[1]{%
 \ifthenelse{\isundefined{\ckboldon}}{#1}{ \textbf{#1} }
}

\begin{document}
\date{}
\title{A Markov-Switching Model Approach to Heart Sound Segmentation and Classification}
\author{
Fuad Noman\footnote{Center for Biomedical Engineering, Faculty of Biosciences and Medical Engineering, Universiti Teknologi Malaysia, 81310 Skudai, Johor, Malaysia (e-mail: mnfuad3@live.utm.my; hussain@fke.utm.my)},
Sh-Hussain~Salleh\textsuperscript{*}, 
Chee-Ming Ting\footnote{Center for Biomedical Engineering, Faculty of Biosciences and Medical Engineering, Universiti Teknologi Malaysia, 81310 Skudai, Johor, Malaysia, and also the Statistics Program, King Abdullah University of Science and Technology, Thuwal, 23955-6900, Saudi Arabia}, 
S. Balqis Samdin\footnote{Statistics Program, King Abdullah University of Science and Technology, Thuwal, 23955-6900, Saudi Arabia.}  
\hspace{0.05cm}, Hernando Ombao\textsuperscript{$\ddagger$} and \\
Hadri Hussain\textsuperscript{*}}

\markboth{IEEE Journal of Biomedical and Health Informatics}%
{Shell \MakeLowercase{\textit{et al.}}: Bare Demo of IEEEtran.cls for Journals}

\maketitle

\begin{abstract}

\textit{Objective:} This paper considers challenges in developing algorithms for accurate segmentation and classification of heart sound (HS) signals. \textit{Methods:} We propose an approach based on Markov switching autoregressive model (MSAR) to segmenting the HS into four fundamental components each with distinct second-order structure. The identified boundaries are then utilized for automated classification of pathological HS using the continuous density hidden Markov model (CD-HMM). The MSAR formulated in a state-space form is able to capture simultaneously both the continuous hidden dynamics in HS, and the regime switching in the dynamics using a discrete Markov chain. This overcomes the limitation of HMM which uses a single-layer of discrete states. We introduce three schemes for model estimation: (1.) switching Kalman filter (SKF); (2.) refined SKF; (3.) fusion of SKF and the duration-dependent Viterbi algorithm (SKF-Viterbi). \textit{Results:} The proposed methods are evaluated on Physionet/CinC Challenge 2016 database. The SKF-Viterbi significantly outperforms SKF by improvement of segmentation accuracy from 71\% to 84.2\%. The use of CD-HMM as a classifier and Mel-frequency cepstral coefficients (MFCCs) as features can characterize not only the normal and abnormal morphologies of HS signals but also morphologies considered as unclassifiable (denoted as X-Factor). It gives classification rates with best gross $F1$ score of 90.19 (without X-Factor) and 82.7 (with X-Factor) for abnormal beats.
\textit{Conclusion:} The proposed MSAR approach for automatic localization and detection of pathological HS shows a noticeable performance on large HS dataset. \textit{Significance:} It has potential applications in heart monitoring systems to assist cardiologists for pre-screening of heart pathologies.

\end{abstract}

\vspace{5mm}

\vspace{-0.05in}

\section{Introduction}

Cardiac auscultation is a critical stage in the diagnosis and examination of heart functionality. Phonocardiogram (PCG) provides a recording of subaudible sounds and murmurs from the heart and allows cardiologists to interpret the closure of the heart valves. Heart sounds can reflect the hemodynamical processes of the heart and provide important screening indications of disease in early evaluation stages. The PCG has been proven as an effective tool to reveal several pathological heart defects such as arrhythmias, valve disease, and heart failure \cite{Liu2016}. The goal of this paper is to develop an automatic method for heart sounds analysis, particularly the segmentation and classification of fundamental heart sounds, which is useful to detect heart pathology in clinical applications.

Several automatic methods for heart sound segmentation have been proposed in the literature. Three main problems must be tackled jointly towards fully automatic heart sound analysis. The first is to detect noise to identify the non-cardiac sounds. The second is to segment heart sounds to localize the main sound components. The third is to classify heart sounds into healthy and pathological classes. The performance of the heart sound segmentation is highly dependent on the preprocessing step. This is relatively simple in noise-free recordings. However, in clinical environments, this is difficult due to both endogenous or exogenous in-band noise sources that overlap with the heart sounds frequency range \cite{Kumar2011}. Accurate localization of the fundamental heart sounds will lead to a more accurate classification of any pathology in systolic or diastolic regions \cite{Springer2016, Springer2014}. 

The heart sound segmentation methods proposed in the literature can be categorized into three groups: the first is the envelope based methods \cite{Liang1997, Huiying1997, Moukadem2013, Sun2014, Choi2008, Yan2010, Ari2008}; the second is feature based methods \cite{Naseri2013, Kumar2006, Varghees2014, Pedrosa2014, Nigam2005, Vepa2008, Papadaniil2014, Arash2011}; the third is machine learning based methods \cite{Oskiper2002, Sepehri2010, Chen2010, Gupta2007, HongTang2010, Rajan2006}, further reviews and details of these methods can be found in \cite{Liu2016, Springer2016}.
Machine learning methods based on probabilistic models show an improved accuracy on heart sound segmentation. Gamero and Watrous \cite{Gamero2003} proposed a hidden Markov model (HMM) approach to detect the S1 and S2 sounds. They used a topology combining two separate HMMs to model the mel-frequency cepstral coefficients (MFCC) of the systolic and diastolic intervals, respectively. The method was evaluated on 80 mostly healthy subjects and achieved a sensitivity of 95\% and positive predictivity of 97\%. Ricke \textit{et al.} \cite{Ricke2005} extended the conventional HMM to a variable-state embedded HMMs method to model the heart sound components (S1, Systole, S2, and Diastole) along with time-variant MFCC, Shannon energy, and regression coefficients. Evaluation only on 9 subjects shows an accuracy of 98\% using eight-fold cross-validation. Gill \textit{et al.} \cite{Gill2005} suggested a modified HMM to allow for a smooth transition between states. 
On 44 heart sound recordings from 17 subjects, the method showed a sensitivity and positive predictivity of 98.6\% and 96.9\% for S1, and 98.3\% and 96.5\% for S2 sound detection. Sedighian \textit{et al.} \cite{Sedighian2014} also used a homomorphic filtering approach to extract envelograms from the heart sound recordings. Envelope peak detection method was used along with two-states HMM to identify the S1 and S2 sound. The method was evaluated on the PASCAL database \cite{Bentley2011} and obtained an average accuracy of 92.4\% for S1 and 93.5\% for S2 sound segmentation. Shmidt \textit{et al.} \cite{Schmidt2010} proposed a duration-dependent HMM method to model the transition duration of each HMM state.
The performance was evaluated on 113 subjects (40 for the training set and 73 for the testing set), the results obtained on the unseen test set were 98.8\% sensitivity and 98.6 positive predictivities. Springer \textit{et al.} \cite{Springer2016} extended the work of \cite{Schmidt2010} by using the hidden semi-Markov model (HSMM) with the modified Viterbi algorithm to detect the beginning and end state of the heart sound signal. 
The method was evaluated on larger heart sound recordings, 10,172 seconds of heart sound collected from 112 (healthy and pathological) subjects admitted to the Massachusetts General Hospital for cardiac screening or in-home recordings including patients with mitral valve prolapse (MVP). The data was split equally into train and test sets. The method obtained an average F1 score of 95.63\% on the unseen test dataset. 
Despite the noticeable performance in identifying heart sounds pathologies, many of the above-mentioned methods were only evaluated on relatively small datasets and mostly from a single source. In contrast, our proposed method will be evaluated on a large standard database. Another major advantage of our approach to heart sounds segmentation is that it is based on modeling of the raw heart sound signals directly, and thus does not require any preliminary stage of feature extraction.

Switching linear dynamic systems (SLDS) \cite{Shumway1991, Ghahramani2000} has been introduced as a generalization of HMM and state space model (SSM). SLDS is capable of modeling changes in time series with a mixture of distinct underlying dynamics which reoccur at certain time intervals.
Most real-world processes are not discrete or exhibit purely linear dynamics. The SLDS is a non-linear model that iteratively segments the data into piecewise stationary regimes by switching between a set of approximately linear dynamic models \cite{Fox2009}. SLDS is widely used in many domains of applications including financial time series \cite{Hamilton1989, Carvalho2007}; motion tracking \cite{Oh2008, Pavlovic2000, Fox2007, XRong2005}; anomaly detection \cite{Ghahramani2000, Oster2015, Montazeri2015, Melnyk2016}; environment \cite{Monbet2017}.
Oster \textit{et al.} \cite{Oster2015} introduced the use of a switching Kalman filter (SKF) for ventricular beat detection in electrocardiogram (ECG) signals. Nasim \textit{et al.} \cite{Montazeri2015} also proposed SKF-based methods with two different switching schemes for apnea bradycardia detection in ECG signals, which showed better performance than a conventional HMM. Samdin \textit{et al.} \cite{Samdin2017} employed a Markov-switching vector autoregressive (MS-VAR) model formulated into a SLDS form to track the state-related changes in functional magnetic resonance imaging (fMRI) and epileptic electroencephalogram (EEG) signals. The approach is able to automatically segment the directed connectivity structure in the multivariate signals into a finite number of reoccurring quasi-stable states. Heart sound signal components exhibit distinct dynamics in the autocorrelation structure at different time intervals, which can be well-captured by a switching autoregressive (AR) process.

In this paper, we develop a unified framework based on Markov-switching AR (MSAR) models with enhanced state inference algorithms to segment the fundamental components of heart sound for subsequent use in classification of heart pathologies. To characterize dynamic cardiac events, we use MSAR models with four states each associated with one of heart sound components. Conventional HMM is less effective when used to segment the raw heart sound signals corrupted by various noise sources (with low signal-to-noise ratio) typically present in the clinical environment. To overcome this limitation, we develop a SLDS formulation by specifying the MSAR as an unobserved latent process to capture the underlying time-variant autocorrelations, and the measured heart sound signals as a contaminated version of this latent process to accommodate the noise effects. To the best of our knowledge, this is the first to apply a MSAR-SLDS for heart sound segmentation.
We introduce two approaches to sequentially infer the latent states of heart sound components. The first is inspired by \cite{Samdin2017} which uses the forward-backward Kalman filter recursions to estimate and smooth the state transition probabilities. 
This approach imposed a constraint on the Markovian transition matrix to form a left-to-right non-ergodic Markov chain allowing only certain pre-specified state transitions according to the temporal order of the heart sound components; The second approach incorporates the Viterbi algorithm to replace the backward-Kalman smoother. In addition to the constrained transition matrix, this approach allows the self-transitions and ensures that mode changes to another state at a certain limit of duration, which corresponds to the durations of each major component in a heart cycle. 

We further employed a continuous-density HMM with Gaussian mixtures for heart sound classification, using the SKF-derived heart-sound segments in the model training.
The Mel-frequency cepstral coefficients (MFCC's) method which widely used in speech analysis was adopted in this paper to extract acoustic features from the heart sound signals. The MFCC is able to represent the frequency contents of the heart sounds in a quasi-logarithmic manner, mimicking the human auditory system. The extracted sequences of MFCC features were computed over sliding windows from each heartbeat. The MFCC features were then modeled using a Gaussian mixture-based HMM approach which shows an improved heart sound classification performance. We consider classification of heart sound classes into three main classes: normal, abnormal and unsure (noisy or X-Factor)\cite{Liu2016}. Incorporating X-Factor class allows the technique to detect the unknown or unclassifiable heart events and reduce the classification of false alarms. In HMM model estimation, each heart sound segment is clustered into four states with 16-Gaussian mixtures, the standard Viterbi algorithm is used to obtain the state sequence, the HMM parameters are then iteratively re-estimated using the expectation-maximization algorithm. The segmentation and classification performance of the proposed method was evaluated under various experimental conditions.

A preliminary version of this work on the segmentation has been reported in \cite{noman2017}. This paper provides a significant extension by presenting a novel, unified framework for both segmentation and classification of heart sounds based on the Markov-switching approach with thorough experimental evaluation on a large database.

\section{Materials and Methods}

\subsection{Heart Sound Database}

An open access heart-sound database which recently published and available online in Physionet/Computing in Cardiology (CinC) Challenge 2016 was used in this study to evaluate the proposed segmentation method \cite{Liu2016}. The database as depicted in Table \ref{Table:table1}, consists of six datasets (\textit{a}  through \textit{f}), collected from different sources by different research groups in both clinical and nonclinical environments \cite{GariD2016}. The database consists of 764 subjects, manually labeled by experts into three classes (2302 normal; 572 abnormal; and 279 unsure), giving a total of 3153 heart sound recordings. The data were recorded at 2000Hz using heterogenous equipment from the four common locations on chest area (aortic, pulmonary, tricuspid, and mitral) with a variety of durations lasting from 5.3s to 122s, 19 hours and 73 minutes in total. Table \ref{Table:table1} summarizes the number of complete heart-beat segments in the dataset, where each segment begins at the start of $S1$ sound until the start of the next $S1$ sound, giving a total of 81498 beats (with 65152 normal and 16346 abnormal segments). 

\begin{table}[!t]
\renewcommand{\arraystretch}{1.3}
\caption{Distribution of complete heart-beat segments in Physionet database.}
\vspace{0.1 cm}
\label{Table:table1}
\centering
\resizebox{0.4\textheight}{!}{
\begin{tabular}{m{1cm} cccc}
  \hline \hline
\multirow{2}{*}{Dataset} & \multicolumn{2}{c}{Beat count} & \multirow{2}{*}{Total beats} & \multirow{2}{*}{Ignored rec.$^\dagger$} \\
\cline{2-3} & Normal & Abnormal &  \\
\hline
 Ds-\textit{a} & 4301  & 9860  & 14161 & 17 \\
 Ds-\textit{b} & 2396  & 589   & 2985  & 122 \\
 Ds-\textit{c} & 356   & 1425  & 1781  & 4 \\
 Ds-\textit{d} & 308   & 493   & 801   & 3 \\
 Ds-\textit{e} & 54783 & 2841  & 57624 & 129$^\ddagger$\\
 Ds-\textit{f} & 3008  & 1138  & 4146  & 6$^\star$ \\
 \hline
 Total         & 65152 & 16346 & 81498 & 281\\
\hline \hline

\multicolumn{4}{m{7cm}}{Those recordings are labeled as noise $^\dagger$
Including recording (e00210)$^\ddagger \hskip 2cm$ 
Including recording (f0043)$^\star $}
\end{tabular}}
\vspace{-0.1in}
\end{table}

The recordings labeled as all--noises were discarded from the segmentation analysis, the remaining recordings were split into train and test datasets with each dataset containing approximately the same number of recordings and heartbeat segments. Table \ref{Table:table2} shows the breakdown of each dataset by heartbeat type (normal or abnormal), this split of the data was chosen to balance the train-test subsets for the performance evaluation of the proposed segmentation and classification methods.

\begin{table}[!t]
\renewcommand{\arraystretch}{1.3}
\caption{Distribution of the Train and Test sets (Segments and Recordings).}
\vspace{0.1 cm}
\label{Table:table2}
\centering
\resizebox{0.6\textwidth}{!}{
\begin{tabular}{m{1cm} cccc|cccc}
\hline \hline
\multirow{3}{*}{Dataset} & \multicolumn{4}{c}{Heart Beats} & \multicolumn{4}{c}{Recordings} \\
\cline{2-9}
\						 & \multicolumn{2}{c}{Normal} & \multicolumn{2}{c}{Abnormal} & \multicolumn{2}{|c}{Normal} & \multicolumn{2}{c}{Abnormal} \\

\cline{2-9} 
     & Train & Test & Train & Test & Train & Test & Train & Test\\
\hline
 Ds-\textit{a} & 2148  & 2153  & 4932 & 4928 & 59   & 57   & 139 & 137\\
 Ds-\textit{b} & 1198  & 1198  & 294  & 295  & 147  & 148  & 36  & 37\\
 Ds-\textit{c} & 177   & 179   & 710  & 715  & 3    & 4    & 10  & 10 \\
 Ds-\textit{d} & 154   & 154   & 246  & 247  & 14   & 12   & 14  & 12\\
 Ds-\textit{e} & 27392 & 27391 & 1420 & 1421 & 889  & 890  & 74  & 72\\
 Ds-\textit{f} & 1502  & 1506  & 568  & 570  & 38   & 39   & 15  & 16\\
 \hline
 Total         & 32571 & 32581 & 8170 & 8176 & 1150 & 1150 & 288 & 284\\
\hline \hline
\end{tabular}}
\vspace{-0.1in}
\end{table}

\subsection{Heart Sound Segmentation}
\begin{figure*}[!th]
\vspace{-0.05in}
\captionsetup[subfigure]{labelformat=empty}
\centering
	\begin{minipage}[t]{1\linewidth}
		\subfigure[]{\includegraphics[width=1\linewidth,keepaspectratio]{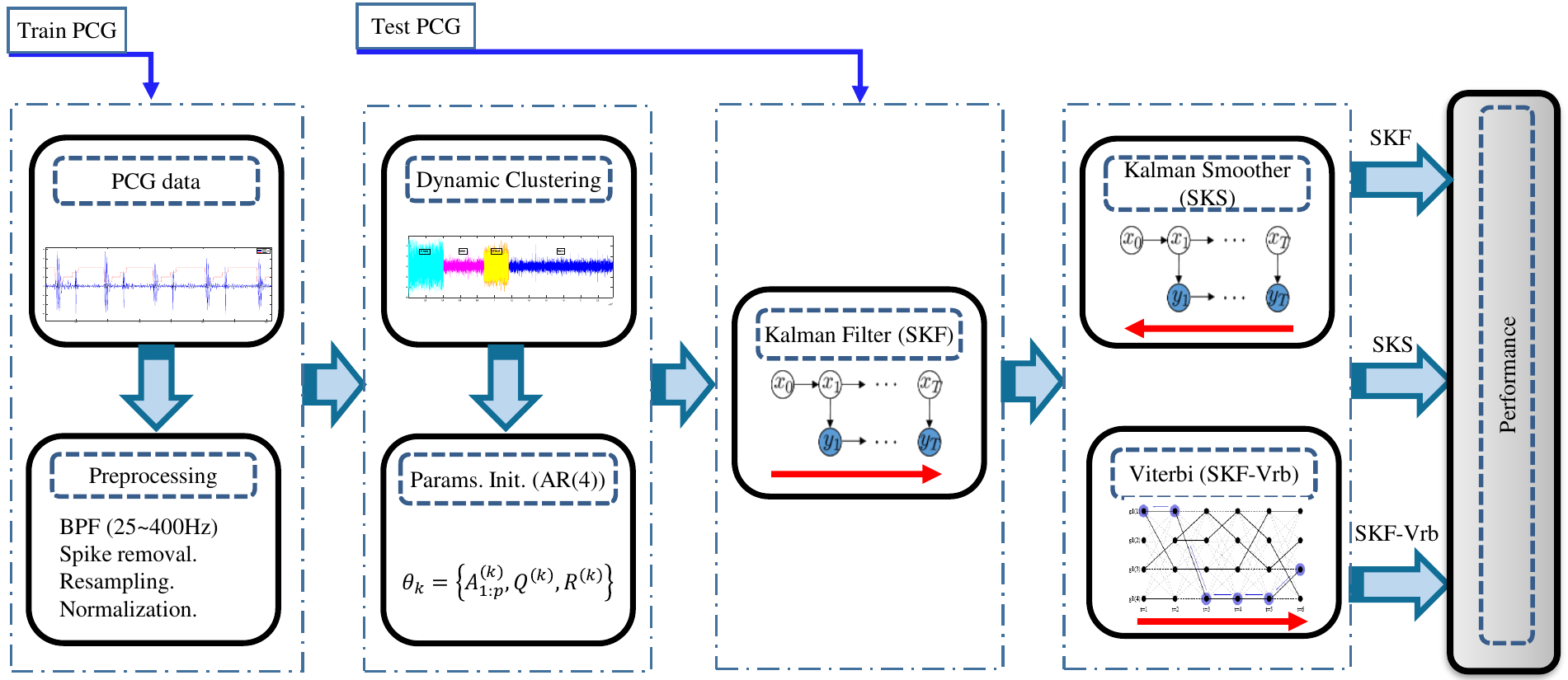}}
	\end{minipage}
	\vspace{-0.4 cm}
\caption{The proposed MSAR-based framework for heart sound segmentation.}
\label{Fig:fig1}
\vspace{-0.1in}
\end{figure*}

Figure \ref{Fig:fig1} shows the proposed framework for heart sound segmentation. The procedure consists of five steps: (1.) Pre-processing to assess the signal quality and filter out the redundant frequency bands (Section B.2). (2.) Dynamic clustering using the reference data labels. (3.) Model parameters initialization. (4.) Switching Kalman filter (SKF) to compute (estimate) the observation likelihood. (5.) Approximate inference algorithms (switching Kalman smoother (SKS) and Viterbi) to estimate the most likely state sequence.

\vspace{-0.14in}
\noindent
\vspace{-0.05in}

\subsubsection{Pre-processing}
However, the recordings labeled with low-quality index were discarded \cite{Liu2016}, different noise sources are still marginally represented in the database. Hence signals were filtered using a Butterworth band-pass filter with cut-off frequencies of 25Hz and 400Hz. The noise spikes were identified and removed using a windowed-outlier filter \cite{Schmidt2010}.  Each recording in the database was shifted and scaled prior to analysis, by subtracting the mean and dividing by standard deviation \cite{Springer2016}.

\subsubsection{Markov-Switching Autoregression (MSAR)}
Modeling the heart sound signal is very challenging because it is nonstationary, nonlinear and periodic time series which consist of repeated heartbeats. Moreover, the clean heart sounds are embedded in various physiological noises and artifacts with a very low SNR. Let ${\bf y}=[\mathrm {y}_1 \ldots,\mathrm{y}_T]'$ be a vector of heart sound time series of length $T$ for the entire recording. We assume an additive noise model for the measured raw heart sound signals as follows
\begin{equation}
 \mathrm{y}_t = \mathrm{x}_t + {\varepsilon}_{t}
\label{Eqn:eqn1*}
\end{equation}
where ${\varepsilon}_{t}$ is a i.i.d. Gaussian observational noise with zero mean and covariance $R$, $\varepsilon_t \sim{N}(0,R)$. The underlying switching dynamics of the clean heart sound signals are assumed to follow a Markov-switching AR process (MSAR), a collection of stationary AR processes that alternate among themselves over time according to an indicator variable $S_t$
\begin{equation}
\mathrm{x}_t = \sum_{p=1}^{P} \varphi_p^{(S_t)}  \mathrm{x}_{t-p} + \eta_t
\label{Eqn:eqn2}
\end{equation}
where $S_t, t=1,\ldots,T$ is a sequence of time-varying state variables taking values in a discrete space $j=1,\ldots,K$; $\{\varphi_p^{(j)}, p=1,\ldots,P \}$ are the AR coefficients at different lags for state $j$; and $\eta_t \sim{N}(0,q)$ is a white Gaussian noise. We assume $S_t$ to follow a hidden Markov chain with transition matrix $Z=[z_{ij}], 1 \leq i,j \leq K$ where $z_{ij} = P(S_t=j|S_{t-1}=i)$ denotes the probability of transition from state $i$ at time $t-1$ to state $j$ at $t$. Each cardiac cycle of heart sound consists of four fundamental components: S1 sound; systolic interval (Sys); S2 sound; and diastolic interval (Dia). The heart sound components exhibit distinct dynamic patterns during different time periods, where each can be modeled as a piecewise-stationary AR process of the MSAR model (\ref{Eqn:eqn2}). Thus, we assume the number of states or regimes as $K=4$ each corresponding to one of the four components ($j=1$: S1, $j=2$: Sys, $j=3$: S2 and $j=4$: Dia). The switching in autocorrelation structure as captured by the state-specific AR coefficients $\varphi_p^{(S_t)}$ between the components is driven by the changes in latent states $S_t$ which indicate which heart-sound component is active at time point $t$. The segmentation of the heart-sound components can be derived indirectly from the state sequence $S_t$. The topology of the Markov chain is set to constrain the transition from one state (or component) to the other in a strict left-to-right sequential order.

Defining a $P \times 1$ hidden state vector of stacked clean heart sound signals $\mathrm{X}_t = \left[ \mathrm{x}_t, \mathrm{x}_{t-1}, \ldots, \mathrm{x}_{t-P+1} \right]$, we can formulate the MSAR plus noise model defined in (\ref{Eqn:eqn1*})-(\ref{Eqn:eqn2}) in a switching linear-Gaussian SSM
\begin{eqnarray}
\label{Eqn:eqn3*}
\mathrm{X}_t & = & A^{(S_t)} \mathrm{X}_{t-1} + \mathrm{w}_t \\ 
\label{Eqn:eqn4}
\mathrm{y}_t & = & C \mathrm{X}_t + \varepsilon_t 
\end{eqnarray}
In the state equation (\ref{Eqn:eqn3*}), the switching AR($P$) process (\ref{Eqn:eqn2}) is written as an $P$-dimensional switching AR(1), where $\mathrm{w}_t = \left[\eta_t,0,\ldots,0\right]$ is a $P \times 1$ state noise, and $A^{(S_t)}$ is a $P$ matrix of AR coefficients switching according to state variables $S_t$
\[
{A}^{(S_t)} =
\left[
  \begin{array}{ccccc}
    \varphi_1^{(S_t)} & \varphi_2^{(S_t)} & \ldots & \varphi_{P-1}^{(S_t)} & \varphi_P^{(S_t)} \\
    1         & 0       & \ldots   & 0             & 0         \\
    0         & 1       & \ldots   & 0             & 0         \\
	\vdots    &         & \ddots   &               & \vdots    \\
	0         & 0       & \ldots   & 1             & 0         \\
  \end{array}
\right].
\]
In the observation equation (\ref{Eqn:eqn4}), the latent MSAR process is observed under noise $\varepsilon_t$ as the measured heart sound signals $\mathrm{y}_t$ via the $1 \times P$ mapping matrix $C = [1,0,\ldots,0]$. We further assume the observation and state noise as white Gaussian processes, i.e. $\varepsilon_t \sim{N}(0,R^{(S_t)})$ and $\mathrm{w}_t \sim{N}(0,Q^{(S_t)})$ with
\[
{Q}^{(S_t)} =
\left[
  \begin{array}{ccccc}
    q^{(S_t)}      & 0 & \ldots & 0  & 0      \\
    0      & 0 & \ldots & 0  & 0      \\
    0      & 0 & \ldots & 0  & 0      \\
	\vdots &   & \ddots &    & \vdots \\
	0      & 0 & \ldots & 0  & 0      \\
  \end{array}
\right].
\]
The noise covariance matrices $R^{(S_t)}$ and $Q^{(S_t)}$ are allowed to switch according to $S_t$. The MSAR model in a state-space form is now fully specified with the model parameters denoted by $\Theta = \left\{Z,A^{(j)},Q^{(j)},R^{(j)}\right\}, j=1,$ $\ldots,K$. The estimation algorithms for the unknown state sequence $S_t$ and model parameters $\Theta$ are given in the following section.

\subsubsection{Dynamic Clustering and Model Initialization}
\begin{figure}[!t]
\vspace{-0.05in}
\captionsetup[subfigure]{labelformat=empty}
\centering
	\begin{minipage}[t]{0.6\linewidth}
		\subfigure[]{\includegraphics[width=1\linewidth,keepaspectratio]{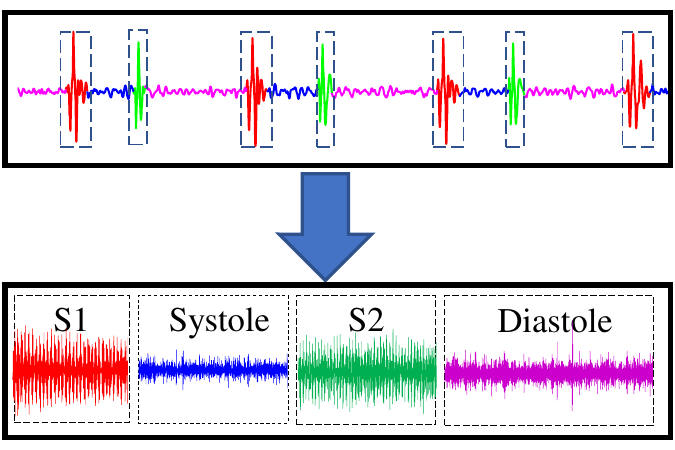}}
	\end{minipage}
	\vspace{-0.4 cm}
\caption{Dynamic clustering of heart sound into four fundamental components.}
\label{Fig:fig3}
\vspace{-0.1in}
\end{figure}

To initialize the MSAR model parameters, we first perform the dynamic clustering to group the heart sound time series data that belongs to the same state or component. This is followed by fitting a separate stationary AR model to the clustered data of each state to obtain the estimators for the state-specific parameters. Conditioned on the known state sequence derived from the expert's manual annotation labels), we partition temporally the time sequence of the heart sound recording in the training set into similar underlying dynamics according to the $K=4$ components. Let ${\bf y}^{(j)}=[\mathrm {y}^{(j)}_1 \ldots,\mathrm{y}^{(j)}_{T_j}]'$ be $T_j \times 1$ vector of concatenated data being clustered to each heart sound component $j=1,\ldots,K$, consisting of the $\mathrm {y}_t$ with $S_t = j$. Figure \ref{Fig:fig3} shows an example of clustering a healthy heart sound signal into four dynamic clusters. Note that the time series data of systoles exhibits the similar dynamic structure as that of the diastole.

Assuming local stationarity for each of these temporal clusters of heart sound signals, we use a simple procedure to initialize the estimates of the MSAR model parameters. Precisely, we assume the concatenated time series of each component to follow a distinct stationary AR($P$) process
\begin{equation}
 \mathrm{y}_t^{(j)} = \sum_{p=1}^{P} {\varphi}^{(j)}_p \mathrm{y}^{(j)}_{t-p} + {\eta}^{(j)}_{t}
\label{Eqn:eqn1}
\end{equation}
We compute the initial estimates of the state-specific AR coefficients $\widehat{\varphi}^{(j)}_p$ by a least-square fitting of the AR($P$) to ${\bf y}^{(j)}$, and the noise variance $\widehat{q}^{(j)}$ based on the estimated residuals $\widehat{\eta}^{(j)}_{t} = \mathrm{y}_t^{(j)} - \sum_{p=1}^{P} \widehat{\varphi}^{(j)}_p \mathrm{y}^{(j)}_{t-p}$ by $\widehat{q}^{(j)} = 1/T_j \sum_{t=1}^{T_j} \left(\widehat{\eta}^{(j)}_{t}\right)^2$. Note that the estimators are initialized based on the manual annotations of the heart sound components, which are subsequently refined based the switching Kalman filter-derived segmentation. The observation noise variance $R$ is also estimated based averaged residuals of the fitted AR over sliding-windowed segments of heart sound signal. The state transition probabilities $z_{ij}$ can be initialized by the frequency of transitions from $S_{t-1} = j$ to $S_t = i$.

\subsubsection{MSAR-based Segmentation Algorithms}
Segmenting the heart-sounds can be cast as the problem of estimating the unknown state sequence $S_t$. Given the sequence of observations $\{\mathrm{y}_t\}_{t=1}^T$, the problem of inference in the switching state-space models is to estimate the posterior probabilities $Pr(S_t=j|\{\mathrm{y}_t\}_{t=1}^T)$ of the hidden state variables $S_t$. 

In this paper, we consider three approaches to estimating the state probabilities given the observation sequence. (1) Switching Kalman filter (SKF) which computes sequentially in a forward recursion the probability densities of the hidden states $P(\mathrm{x}_t |\{\mathrm{y}_t\}_{t=1}^t)$ and $P(S_t |\{\mathrm{y}_t\}_{t=1}^t)$ given observations up to time $t$; (2) Switching Kalman smoother (SKS) (or Rauch-Tung-Streibel smoother RTS) computes in a backward recursion refined estimates of densities $P(\mathrm{x}_t |\{\mathrm{y}_t\}_{t=1}^T)$ and $P(S_t |\{\mathrm{y}_t\}_{t=1}^T)$ given the entire observation sequence of length $T$; (3) Fusion of SKF and extended duration-dependent Viterbi algorithm (SKS-Viterbi) suggested by \cite{Springer2016, Schmidt2010} which decodes the most likely sequence of states given the state probabilities from the one-step ahead Kalman Filter predictions $P(S_t=j | M_{t|t}^j )$

\paragraph{Switching Kalman Filter (SKF):}

Algorithm \ref{Algo:alg3} summarizes the procedure of SKF for estimating the hidden state parameters given the raw heart sound observations $\{\mathrm{y}_t\}_{t=1}^T$ and estimated model parameters for each state $\widehat{\Theta} = \left\{\widehat{Z},\widehat{A}^{(j)},\widehat{Q}^{(j)},\widehat{R}^{(j)}, j=1,\ldots,K \right\}$. Refer to \cite{Murphy1998} for further details. Given $\widehat{\Theta}$ and initial state probabilities $M_0^j=[1,0,\ldots,0]$, for each time $t$, a run of $K^2$ Kalman filters is performed recursively to compute the mean and covariance of the component filtered densities of $\mathrm{x}_t$ (denoted as $\mathrm{x}_{t|t}^{ij}$ and $P_{t|t}^{ij}$) for all pairs $(i,j)$ and the corresponding likelihood function $L_t^{ij}$. The filtered state probability of $S_t$ can be defined by
\vspace{-0.02in}
\begin{eqnarray}
M_{t|t}^j & = & P(S_t=j | \{\mathrm{y}_t\}_{t=1}^t) \notag \\
& = & \sum_i M_{t-1,t|t}^{ij} \label{Eqn:eqn8}
\end{eqnarray}

where $M_{t-1,t|t}^{i,j}= P(S_{t-1}=i,S_t=j | \{\mathrm{y}_t\}_{t=1}^t)$ is computed from the $M_{t-1|t-1}^i$ at previous time $t-1$ weighted by the likelihood $L_t^{ij}$ and the transition probabilities $z_{ij}$ as follows
\begin{equation*}
M_{t-1,t|t}^{ij} = \frac{L_t^{ij} z_{ij} M_{t-1|t-1}^i}{\sum_i \sum_j L_t^{ij} z_{ij} M_{t-1|t-1}^i}  \label{Eqn:eqn8}
\end{equation*}
After the filtering at each time $t$, the component densities ($\mathrm{x}_{t|t}^{ij}$ and $P_{t|t}^{ij}$) weighted by $W_t^{i|j} = M_{t-1,t|t}^{ij}/M_{t|t}^j$ are collapsed to give the mean and covariance of the filtered densities ($\mathrm{x}_{t|t}^{j}$ and $P_{t|t}^{j}$).

\begin{algorithm}[!t]
\noindent\textbf{Inputs}: $\mathrm{x}_0^{ij}, P_0^{ij}, M_0^j, \{\mathrm{y}_t\}_{t=1}^T, A, C, R, Q, Z$\\
\noindent\textbf{Outputs}: $M_{t|t}^j$,$\mathrm{x}_{t|t}^j$, $P_{t|t}^j$

\vspace{-0.08in}
\noindent\hrulefill
\begin{algorithmic}[1]
\For {$t = 1, 2, \ldots, T$}
\For {$j = 1, \dots , K$}
\For {$i = 1, \dots , K$}
\State {$[\mathrm{x}_{t|t}^{ij}, P_{t|t}^{ij},L_t^{ij}] = $ Filter$(\mathrm{x}_{t-1|t-1}^i,P_{t-1|t-1}^i,$\\ \hspace{12em} $ A^j, C, Q^j, R^j)$}
\EndFor
\EndFor
\For {$j = 1, \dots , K$}
\State {$[M_{t|t}^j,W_t^{i|j}] = $ FilterProbs$(L_t^{ij},Z^{ij}, M_{t-1|t-1}^i)$}
\State {$[\mathrm{x}_{t|t}^j, P_{t|t}^j]$ = Collapse$(\mathrm{x}_{t|t}^{ij},P_{t|t}^{ij},W_t^{i|j})$}
\EndFor
\EndFor
\end{algorithmic}
\caption{: Switching Kalman filter}
\label{Algo:alg3}
\end{algorithm}

\paragraph{Switching Kalman Smoother (SKS):}

Algorithm \ref{Algo:alg4} summarizes the procedure of SKS. In a backward recursion, a mixture of $K^2$ Kalman smoothers is run to compute component smoothed densities of $\mathrm{x}_t$ for all pairs $(j,k)$ (with mean $\mathrm{x}_{t|T}^{jk}$ and covariance $P_{t|T}^{jk}$) given the entire observation $\{\mathrm{y}_t\}_{t=1}^T$ based on the filtered densities computed in the SKF. The smoother state probability of $S_t$ is defined as
\vspace{-0.02in}
\begin{eqnarray}
M_{t|T}^j & = & P(S_t=j | \{\mathrm{y}_t\}_{t=1}^T) \notag \\
& = & \sum_k M_{t,t+1|T}^{jk} \label{Eqn:smoothSt}
\end{eqnarray}
where $M_{t,t+1|T}^{jk} = P(S_{t}=j,S_{t+1}=k | \{\mathrm{y}_t\}_{t=1}^T)$ can be computed based on the filtered state probabilities $M_{t|t}^j$ and the smoothed probabilities $M_{t+1|T}^k$ at $t+1$ as follows
\begin{equation*}
M_{t,t+1|T}^{jk} = \frac{M_{t|t}^j z_{jk}}{\sum_j' M_{t|t}^{j'} z_{j'k}} M_{t+1|T}^k
\end{equation*}
Finally, the component densities ($\mathrm{x}_{t|T}^{jk}$ and $P_{t|T}^{jk}$) weighted by $W_t^{k|j} = M_{t,t+1|T}^{jk}/M_{t|T}^j$ are collapsed to give the mean and covariance of the smoothed densities ($\mathrm{x}_{t|T}^{j}$ and $P_{t|T}^{j}$).

\begin{algorithm}[!t]
\noindent\textbf{Inputs}: $\{\mathrm{y}_t\}_{t=1}^T, A, R, Q, Z, \mathrm{x}_{t|t}^j, P_{t|t}^j, M_{t|t}^j$ \\
\noindent\textbf{Outputs}: $M_{t|T}^j$, $\mathrm{x}_{t|T}^j$, $P_{t|T}^j$

\vspace{-0.08in}
\noindent\hrulefill
\begin{algorithmic}[1]
\For {$t = T, T-1, \ldots, 1$}
\For {$j = 1, \dots , K$}
\For {$k = 1, \dots , K$}
\State {$[\mathrm{x}_{t|T}^{jk}, P_{t|T}^{jk}] = $ Smooth$(\mathrm{x}_{t+1|T}^k,P_{t+1|T}^k,\mathrm{x}_{t|t}^j,$\\ \hspace{12em} $ P_{t|t}^j, A^k, Q^k, Z^{jk})$}
\EndFor
\EndFor
\For {$j = 1, \dots , K$}
\State {$[M_{t|T}^j,W_t^{k|j}] = $ SmoothProbs$(M_{t|t}^j, M_{t+1|T}^k)$}
\State {$[\mathrm{x}_t^j, P_t^j]$ = Collapse$(\mathrm{x}_t^{jk},P_t^{jk},W_t^{k|j})$}
\EndFor
\EndFor
\end{algorithmic}
\caption{: Switching Kalman Smoother}
\label{Algo:alg4}
\end{algorithm}

\paragraph{SKF with Viterbi Algorithm:}

Under the Markovian assumption of the standard SKF, the sojourn time or dwell time (the number of consecutive time points spent in a specific state before transitioning to other states) is geometrically distributed, i.e., the probability of remaining in a state decreases as the sojourn time increases. This tends to induce unrealistically fast switching states and may not be appropriate for stationary processes such as each heart sound component with possibly long period of time in the same regime. To overcome this limitation, we introduce a two-step procedure by combining the SKF with the duration-dependent Viterbi algorithm which was first introduced by \cite{Schmidt2010} and extended in \cite{Springer2016}. 
The duration-dependent Viterbi algorithm incorporates explicitly the information about each state expected duration (i.e. heart rate \textemdash HR, systolic interval \textemdash tSys) which are estimated from the testing heart sound recording using autocorrelation analysis. The duration probabilities $dP$ are estimated from the data for each of the four heart sound states.

With an initialized $\delta_1^j$, the algorithm computes the state probability in a forward recursion

\vspace{-0.2in}
\begin{center}
\begin{equation}
\delta_t^j = \max_{d}\Biggl[\max_{i\neq{j}} \quad [\delta_{t-d}^i a_{ij}]\quad  {dP}_d^j \quad {\displaystyle \prod_{s=0}^{d-1}} \{M^j_{t|t}\}_{t=t-s} \Biggr]
\label{Eqn:eqn9}
\end{equation}
\end{center}

\noindent
for $1\leq t \leq T$, $1\leq i,j \leq K$, ${dP}_d^j$ is the duration probabilities for state $j$ for $1\leq d \leq d_{max}$ with $d_{max}$ the number of time points for each heartbeat with reference to the estimated heart rate. Note that we incorporate the SKF state probability $M_{t|t}^j = P(S_t=j | \{\mathrm{y}_t\}_{t=1}^t) \propto P(\{\mathrm{y}_t\}_{t=1}^t| S_t=j)P(S_t=j)$ which takes into account the observations up to time $t$ instead of only the current observation $P(\mathrm{y}_t| S_t=j)$ in the original duration-dependent Viterbi algorithm.
 The state duration argument and the state sequence that maximize (\ref{Eqn:eqn9}) are stored in $D_t^j$ and $\psi_t^j$ respectively. The most likely state sequence is obtained stored in $\psi_t^j$, $\psi_t^j = \underset{1\leq i \leq K}{\mathrm{argmax}} [\delta_{t-D_t^j}^i a_{ij}]$.

The psuedocode of the extended Viterbi algorithm is shown in Algorithm (\ref{Algo:alg5}). Refer \cite{Springer2016} for more details. In Algorithm (\ref{Algo:alg5}), the $\delta_t^j$ is the highest state probability for each state $j$ at time $t$ for all duration probabilities ${dP}_d^j$  from $1$ to $d_{max}$. the state probabilities are updated only if current $\delta_t^i$ is higher than the $\delta_{t-1}^i$ in the processing window $1$ to $d_{max}$. The back-tracking procedure is initialized by finding the maximum probability of $\delta_t^i$ in the interval $T:T+d_{max}-1$ after the end of actual signal. The state index that maximizes $\delta_{T*}^i$ is stored in $q_{T*}^*= \mathrm{argmax}_i [\delta_t^i]$. The optimal path $q_t^*$ is obtained by back-tracking $\psi_T^{q_t^*}$ and $D_T^{q_t^*}$ such that $q_{t-d^*-1}^* = \psi_{q_t^*}$, where $t=T-1,\ldots,1$. 

\begin{algorithm}[!t]
\noindent\textbf{Inputs}: initials $\pi_0,HR,tSys$\\
\noindent\textbf{Outputs}: $q_t$.

\vspace{-0.12in}
\noindent\hrulefill
\begin{algorithmic}[1]
\State {$[\{M_t^j\}_{t=1}^T]=$ SKF$(\{\mathrm{y}_t\}_{t=1}^T), A, R, Q, Z, \mathrm{x}_0, P_0, M_0^j)$}
\State {Initialization: $[a_{ij}, \delta_1^j,d_{max}] = $(\textit{HR}, $tSys, \{M^j_{t|t}\}_{t=1}, \pi_0)$}
\For {$t = 2 : T+d_{max}-1$}
\For {$i,j = 1 : K$}
\For {$d = 1 : d_{max}$}
\State {$w_s = t-d,\quad 1\leq w_s \leq T-1$}
\State {$w_e = t,\quad 2\leq w_e \leq T$}
\State {$\delta_t^j = \max_{d}\Bigl[\max_{i\neq{j}} [\delta_{w_s}^i a_{ij}]\hspace{0.3em} . \hspace{0.3em}{dP}_d^j \hspace{0.3em} . \hspace{0.3em}$ \\ \hspace{10em}${\prod_{s=w_s}^{w_e}} \{M^j_{t|t}\}_{t=s} \Bigr]$}
\State {$D_t^j = \arg \max_{d}\Bigl[\max_{i\neq{j}} [\delta_{w_s}^i a_{ij}]\hspace{0.3em} . \hspace{0.3em}{dP}_d^j \hspace{0.3em} . \hspace{0.3em}$ \\ \hspace{12em}${\prod_{s=w_s}^{w_e}} \{M^j_{t|t}\}_{t=s} \Bigr]$}
\State {$\psi_t^j = \arg \max_{1\leq i \leq K} [\delta_{t-D_t^j}^i a_{ij}]$}
\EndFor
\EndFor
\EndFor
\noindent
\State {$T* = \arg \max_{t}[\{\delta_t^i\}_{t=T}^{T+d_{max}-1}] \qquad 1 \leq i \leq K$}
\State {$q_{T*}^*= \arg \max_i[\delta_{T*}^i]$}
\State {$t=T*$}
\While {$t > 1$} $//$Backward Viterbi procedure
\State {$d^* = D_t^{q_t^*}$}
\State {$\{q\}_{t-d^*}^{t-1}=q_t^*$}
\State {$q_{t-d^*-1}^*=\psi_t^{q_t^*}$}
\State {$t = t-d^*$}
\EndWhile
\end{algorithmic}
\caption{: SKF-Viterbi Algorithm.}
\label{Algo:alg5}
\end{algorithm}

\subsection{Heart Sound Classification}

\begin{figure}[!t]
	\begin{minipage}[t]{0.7\linewidth}
		\centering
		\subfigure[\label{Fig:fig4a}]{\includegraphics[width=1\linewidth,keepaspectratio]{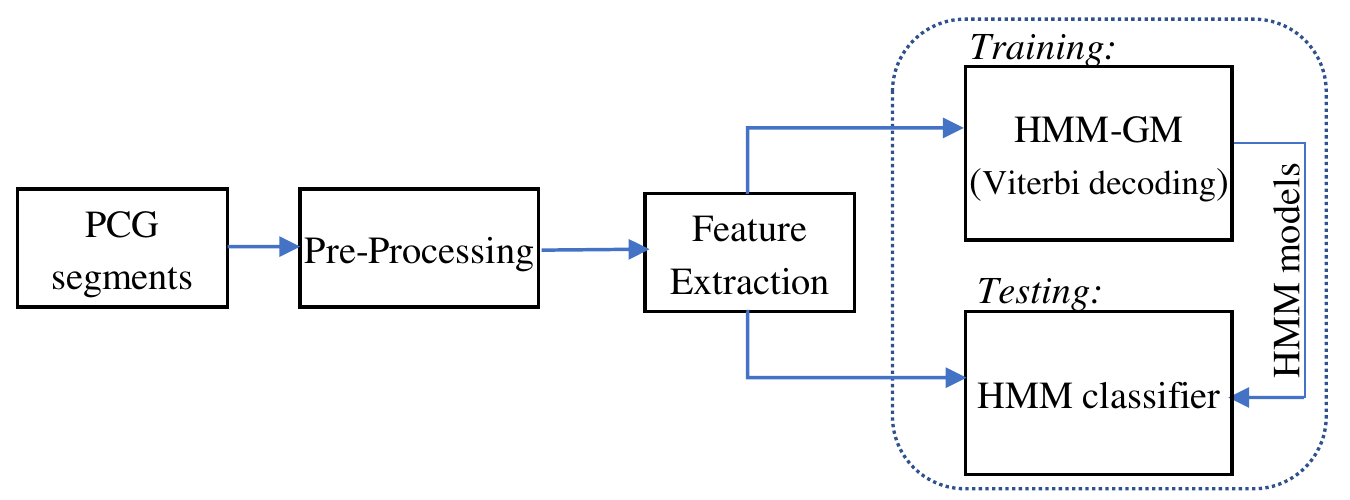}}
	\end{minipage}
	\centering
	\begin{minipage}[t]{0.7\linewidth}
		\centering
		\subfigure[\label{Fig:fig4b}]{\includegraphics[width=1\linewidth,keepaspectratio]{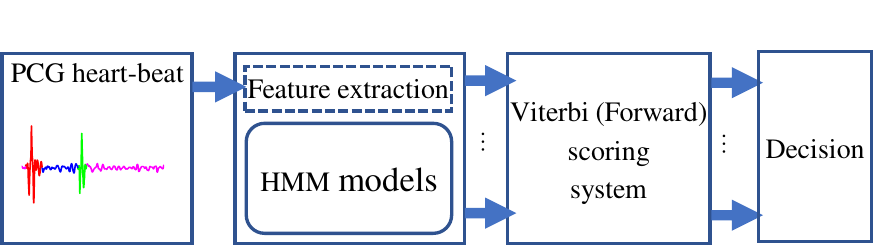}}
	\end{minipage}
	\vspace{-0.2 cm}
\caption{ (a) The overall classification system design for training and testing the HMM models. (b) HMM testing procedure.}
\label{Fig:hmm_train_test}
\vspace{-0.1in}
\end{figure}

In this section, we present an automatic classification of healthy and pathological heart sound recordings using hidden Markov models (HMM) based on the heart-beat segmentation obtained by the switching Kalman filters. The distribution of train and test sets in the database used for evaluation is given in Table \ref{Table:table2}. The heart sound recordings were preprocessed and then segmented using procedures described in Section 2.B, such that each segment covers a complete heart-beat cycle (start of S1 sound to the consequent S1 sound). The Mel-frequency cepstral coefficients (MFCCs) widely used in speech signal processing are adapted for feature extraction.
These MFCC features are then used as input to the HMMs with Gaussian mixture observation density. Figure \ref{Fig:fig4a} illustrates the different steps used in the evaluation of the heart sound classification system.

\subsubsection{Feature Extraction}
A sequence of short-time MFCC feature vectors was extracted from each heart sound segment based on a sliding-window approach using windowed frames of 50ms with 10ms overlap. A Hamming window was used to minimize the discontinuities at the frame edges. For each frame, a set of MFCCs is computed from the short-time spectrum. Each frame was first passed through a first order FIR to spectrally flatten the signal. A discrete Fourier transform (DFT) was applied to each windowed frame and energy at each $mel$ bandwidth (with a value of 20 to 24 in $mel$ scale) was calculated. By taking the logarithm and cosine transform, a vector of 12 MFCCs was derived for each frame.

\subsubsection{HMM Training and Evaluation}

The HMM is a probabilistic model that can capture the dynamical changes of the heart sounds by making inferences about the likelihood of being in certain discrete states. In this paper, a continuous HMM with Gaussian mixtures (GM) consisting of four states (left-to-right, no skipping) and 16 Gaussian mixtures (probability density functions) for each state was used. A set of HMM parameters is denoted by $\lambda =(\boldsymbol{\pi},{\bf A},{\bf B})$ where $\boldsymbol{\pi} = [\pi_{i}]$ with $\pi_{i} = P[q_1=S_i], 1\leq i\leq K$ are the initial state probabilities and ${\bf A} = [a_{ij}]$ is $K \times K$ transition matrix with $a_{ij} = P[q_{t+1} = S_i|q_t=S_j], 1\leq i, j\leq K$. Let $\mathbf{O}_t = [o_{1t}, \ldots, o_{Nt}]^{'}$ be the $N \times 1$ MFCC feature vector at time $t$. The observational emission probability ${\bf B} = \{b_j(x)\}, 1 \leq j \leq K$ at each state $j$ is defined by a Gaussian mixture model
\begin{equation}
b_j(\mathbf{O}_t) = \sum_{m=1}^{M} c_{jm} N(\mathbf{O}_t;\boldsymbol{\mu}_{jm},\boldsymbol{\Sigma}_{jm}), 1 \leq j \leq K
\label{Eqn:eqnHMM1}
\end{equation}
where $\boldsymbol{\mu}_{jm}$ and $\boldsymbol{\Sigma}_{jm}$ are respectively the mean vector and covariance matrix of the $m$-th mixture component with mixture weight $c_{jm}$ at state $j$. Here, we set the number of mixture components as $M=16$ per state.

\paragraph{Training \& Testing:}
The training and testing of the HMMs are illustrated in Fig. \ref{Fig:fig4a} and Fig.\ref{Fig:fig4b}. Given the training observation sequences $\mathbf{O}_1, \ldots, \mathbf{O}_T$ (a complete heart-beat cycle $–$ $S1$, systole, $S2$, diastole), the HMM model parameters were estimated by maximizing the likelihood function. The training of an HMM involves initialization of model parameters followed by iterative re-estimation of the parameters via expectation-maximization algorithm (the Baum-Welch algorithm) until convergence. The segmental K-means algorithm was used in model initialization by first aligning the observations to the corresponding state via the Viterbi algorithm and partitioning the observations into each mixture component by K-means clustering. Separate HMMs were trained for the normal and abnormal heart sounds. Given an unknown testing heart sound segment, the Viterbi algorithm was used to compute the approximate likelihood scores for each HMM model based on the most likely state sequence. The testing heart sound signal will be classified to the model with the highest likelihood score. 

\paragraph{Model evaluation:}
The performances of trained HMM models were evaluated on their ability to correctly classify a given heart sound heartbeat segment within the test set of recordings, into normal or abnormal classes. In order to build the confusion matrix to assess the classification performance, each heartbeat was compared to the existing HMM models. Three different classes were considered in this study, the normal class is denoted by $N$, the abnormal by $A$, and the unsure (X-Factor) class by $Q$. One main motivation of this study is the detection of abnormal heartbeats (or records). We used a large database collected from different sources in different clinical environments where some of the recordings are labeled as noisy or unclassifiable. The proposed approach was evaluated with and without incorporating the noisy (X-Factor) recordings for both heartbeat and recording classification levels. For classification without involving the X-Factor segments or recordings, we used performance metrics as in \cite{Springer2016} such as sensitivity ($Se$), positive productivity ($P_+$), accuracy ($Acc$), and ($F1$) score computed from the confusion matrix.

For classification including the X-Factor class, we used a performance metric proposed by \cite{Liu2016} to compute the overall performance based on the number of beats or recordings classified as normal, abnormal, or X-Factor. The signal quality indices are provided along with the database, Table \ref{Table:table8} illustrates the partitioning of X-Factor recordings into the train and test sets. Total 279 recordings were labeled by cardiologists as unsure (hard to classify) which we consider it as X-Factor recordings in this study.

\begin{table}[!t]
\renewcommand{\arraystretch}{1.3}
\caption{Training and testing sets for X-Factor class.}
\vspace{-0.1 cm}
\label{Table:table8}
\centering
\begin{tabular}{m{1cm} cccccccc}
\hline \hline
	\multirow{3}{*}{Dataset} & \multicolumn{4}{c}{Abnormal} & \multicolumn{4}{c}{Normal}\\
	\cline{2-9}
	 & \multicolumn{2}{c}{Segments} & \multicolumn{2}{c}{Records} & \multicolumn{2}{c}{Segments} & \multicolumn{2}{c}{Records}\\
	 \cline{2-9}
	 & Train & Test & Train & Test & Train & Test & Train & Test \\
	 \hline
	 DS-\textit{a} & 216 & 222 & 8  & 8  & 35   & 0    & 1  & 0 \\
	 DS-\textit{b} & 120 & 125 & 15 & 16 & 360  & 368  & 45 & 46\\
	 DS-\textit{c} & 45  & 91  & 2  & 2  & 0    & 0    & 0  & 0 \\
	 DS-\textit{d} & 12  & 21  & 1  & 1  & 8    & 0    & 1  & 0 \\
	 DS-\textit{e} & 497 & 472 & 18 & 19 & 1045 & 1044 & 45 & 46\\
	 DS-\textit{f} & 32  & 63  & 1  & 2  & 30   & 40   & 1  & 1 \\
	 \hline
	 Total         & 904 & 994 & 45 & 48 & 1478 & 1452 & 93 & 93\\
	 \hline \hline
	\end{tabular}
\vspace{-0.1in}
\end{table}

We computed the modified sensitivity ($Se$), specificity ($Sp$), accuracy ($MAcc$), and $F1$ from the confusion matrix including X-Factor as
 
\vspace{-0.1in}
\begin{equation}
Se = \frac{wa_1 \times Aa_1}{Aa_1 + Aq_1 + An_1} + \frac{wa_2 \times (Aa_2 + Aq_2)}{Aa_2 + Aq_2 + An_2}
\label{Eqn:eqn16}
\end{equation}

\begin{equation}
Sp = \frac{wn_1 \times Nn_1}{Na_1 + Nq_1 + Nn_1} + \frac{wn_2 \times (Nn_2 + Nq_2)}{Na_2 + Nq_2 + Nn_2}
\label{Eqn:eqn17}
\end{equation}

\begin{equation}
MAcc = \frac{Se+Sp}{2}
\label{Eqn:eqn18}
\end{equation}

\noindent
where $wa_{1,2}$ and $wn_{1,2}$ are the percentages of good/poor signal quality in all abnormal and normal recordings (training set) which were used as weights to calculate the $Se$ and $Sp$ respectively. $A$ and $N$ are the true labels of abnormal and normal classes, where the $a$, $q$ and $n$ are the algorithm labels of abnormal, X-Factor and normal classes respectively. For example, $A_{a1,2}$ are the total number of good/poor abnormal (beats or recordings) which were recognized as abnormal class.

We followed \cite{Oster2015} method to calculate the penalized $F1$ score, where a penalty $\alpha$ was applied to $An$ and $Na$ to ensure that all beats that are not considered as belonging to X-Factor. The penalized $F1$ score was computed as follows
\begin{equation}
F1 = \frac{2(\alpha +1)Aa_1}{2(\alpha+1)Aa_1 + \alpha(An_1 + Na_1)+(Aq_1+Nq_1)}
\label{Eqn:eqn19}
\end{equation}
\noindent
where $\alpha = 10$ is the weight or penalty to control the incorrect normal or abnormal classification due to the inclusion of X-Factor class. The $Aq$ beats were considered the pseudo false negative ($PFN$), and $Nq$ the pseudo false positive ($PFP$).

\section{Results and Discussions}

\subsection{Heart Sound Segmentation}

We compare the performance of the three different segmentation algorithms: SKF, SKS, and KF-Viterbi, in annotating the dynamic changes in the heart sound recordings. The performance was evaluated on all recordings in the unseen testing dataset, can be seen in Table \ref{Table:table2} and Table \ref{Table:table8}. The switching Kalman filter algorithms were initialized by fitting a stationary autoregressive model of order ($P=4$) on each state observation sequence in a recording-specific manner. The parameters of the MSAR model were computed by averaging parameter estimates overall recordings in the training dataset.

\vspace{-0.14in}
\noindent
\begin{center}
\begin{figure}[!t]
\vspace{-0.05in}
\captionsetup[subfigure]{labelformat=empty}
\centering
	\begin{minipage}[t]{0.6\linewidth}
		\subfigure[]{\includegraphics[width=1\linewidth,keepaspectratio]{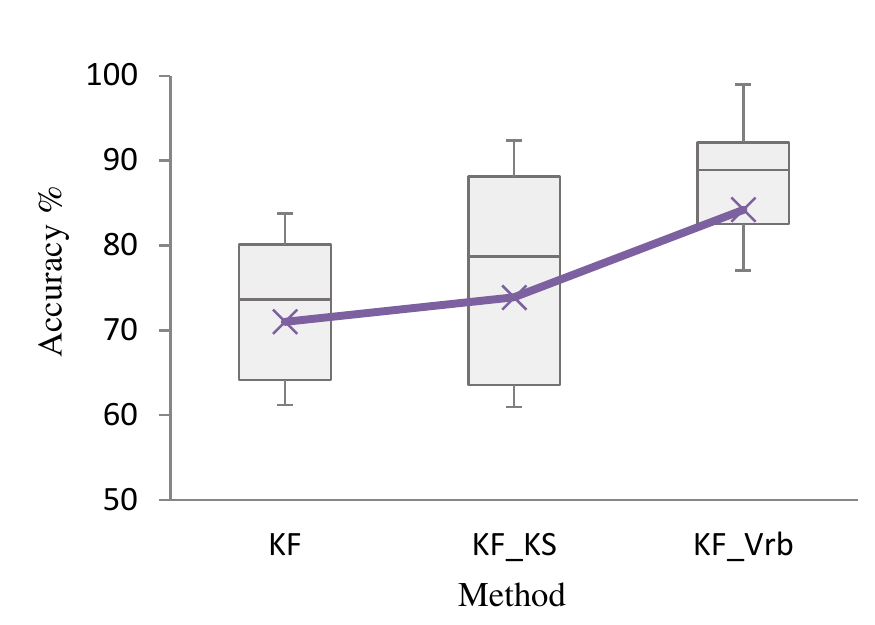}}
	\end{minipage}
	\vspace{-0.8 cm}
\caption{Segmentation performance box-plots using the test dataset (Table \ref{Table:table2}). KF: Kalaman filter segmentation approach, KF\_KS: Kalman Smoother segmentation, KF\_Vrb: fusion of Kalman filter and Viterbi algorithm.}
\label{Fig:fig6b}
\vspace{-0.1in}
\end{figure}
\end{center}
\vspace{-0.2in}

Fig. \ref{Fig:fig6b} shows the results on unseen testing datasets. The models were initialized by fitting the AR(4) on the train dataset dynamic clusters. We can see that the segmentation accuracies of unseen dataset dropped slightly in both SKF and SKS, while the SKF-Viterbi maintained higher performance of 84.2\%. The fusion of SKF and duration-dependent Viterbi algorithm improves the average performance of SKF form 71\% to 84.2\%.

\begin{table}[!t]
\renewcommand{\arraystretch}{1.3}
\caption{Average segmentation performance on selected balanced testing set.}
\vspace{0.1 cm}
\label{Table:table91}
\centering
\resizebox{0.5\textwidth}{!}{
\begin{tabular}{cccccc}
	\hline \hline
	\multirow{2}{*}{Method} & \multirow{2}{*}{Interval} & \multicolumn{4}{c}{\textit{Performance mean $\pm$ SD (\%)}} \\
	\cline{3-6}
	& & $Se$ & $P_+$ & $F1$ & $Acc$\\
	\hline
	\multirow{4}{*}{\begin{tabular} [m{1cm} ]{@{} m{1cm} @{}} SKF \end{tabular}} & \textit{S1} & 74 $\pm$ 12 & 69 $\pm$ 17 & 71 $\pm$ 13 & \multirow{4}{*}{71 $\pm$ 13}\\
	& \textit{Sys} & 61 $\pm$ 21 & 64 $\pm$ 18 & 61 $\pm$ 19 & \\
	& \textit{S2} & 33 $\pm$ 17 & 61 $\pm$ 28 & 40 $\pm$ 20 & \\
	& \textit{Dia} & 85 $\pm$ 12 & 78 $\pm$ 10 & 81 $\pm$ 10 & \\
	\hline
	\multirow{4}{*}{\begin{tabular} [m{1cm} ]{@{} m{1cm} @{}} SKS \end{tabular}} & \textit{S1} & 77 $\pm$ 16 & 74 $\pm$ 20 & 74 $\pm$ 17 & \multirow{4}{*}{74 $\pm$ 18}\\
	& \textit{Sys} & 67 $\pm$ 25 & 68 $\pm$ 23 & 67 $\pm$ 24 & \\
	& \textit{S2} & 55 $\pm$ 24 & 60 $\pm$ 28 & 55 $\pm$ 25 & \\
	& \textit{Dia} & 81 $\pm$ 21 & 83 $\pm$ 14 & 81 $\pm$ 17 & \\
	\hline
	\multirow{4}{*}{\begin{tabular} [m{1cm} ]{@{} m{1cm} @{}} SKF-Viterbi \end{tabular}} & \textit{S1} & 77 $\pm$ 15 & 85 $\pm$ 16 & 81 $\pm$ 15 & \multirow{4}{*}{84 $\pm$ 14}\\
	& \textit{Sys} & 86 $\pm$ 18 & 87 $\pm$ 17 & 81 $\pm$ 17 & \\
	& \textit{S2} & 63 $\pm$ 20 & 76 $\pm$ 21 & 68 $\pm$ 19 & \\
	& \textit{Dia} & 91 $\pm$ 12 & 89 $\pm$ 12 & 90 $\pm$ 12 & \\
	\hline\hline
	\multicolumn{6}{m{10cm}}{\textit{S1}: S1 sound, \textit{Sys}: systolic, \textit{S2}: S2 sound, \textit{Dia}: diastolic 
SD: standard deviation, KF: Kalman filter, KS: Kalman smoother.}
\end{tabular}}
\vspace{-0.2in}
\end{table}

The study presented here investigated new approaches for the segmentation of fundamental heart sounds (S1, Systole, S2, and Diastole) from a single channel heart sound recording without using any reference signals for the labeling process. The results show that using the backward SKS slightly outperforms the SKF method, increasing the accuracy by almost 4\%. However, fusing the duration-dependent Viterbi with the SKF resulted in a significant improvement in heart sound segmentation, achieving almost 10\% higher accuracy.           

The overall performance results of the three proposed approaches on the unseen (not trained) data set, for each fundamental heart sound, are presented in Table \ref{Table:table91}. It is important to note that, the results in this table are calculated with zero tolerance between the ground truth and the estimated labels. The confusion matrix is calculated such that the observation at time $t$ is true positive if it's state matching the ground truth labels, otherwise is considered as false positive. The set of equations provided in \cite{Mariano2011} were used in this paper to calculate the $Se$, $P_+$, $F1$ and global accuracy $Acc$. The Viterbi based approach outperforms both the SKF and SKS achieving global accuracy of 84 $\pm$ 14\% on the hidden testing set, with the highest detection of diastolic intervals.   

The state-of-the-art method \cite{Springer2016} involves a logistic regression model with multivariate normal (MVN) distribution computed from four-dimensional feature vectors extracted from each heart sound recording. The use of such higher dimensional feature space allows the model to adequately best capture the underlying dynamics of the four-state observations. However, the proposed methods in this paper ignore the feature extraction phase and use a down-sampled version of the raw heart sound recordings, in which the Kalman filter infers the state probabilities given a univariate heart sound observation sequence.

\subsection{Heart Sound Classification}

In this section, we evaluate the performance of HMM in abnormal heart sound morphology classification. The proposed technique can perform classification based on beat-level and recording-level paradigms. In the beat-level approach, each heartbeat (segment) was individually classified and assigned to a normal, abnormal, or X-Factor class. Where in recording-level, the classification scores for all heartbeats belonging to the same recording were combined (voting), each recording is classified as abnormal only when the proportion of beats assigned to abnormal class is dominant. The beat-level approach substantially expands the number of training instances, which allows the machine learning application to learn more about the heart sound underlying dynamics for each class. The database provides global (recording-level) labels where each record has been assigned to an abnormal or normal class, we assumed all the beats of a given abnormal recording are also abnormal. Hence, if only a small portion of a recording was corrupted by noise, the recording will not be classified as noisy (X-Factor).

In addition to the beat-level and recording-level classification, two approaches of train-test data partitioning were also investigated to evaluate the performance of the HMM models. The first approach, the whole beats were split into balanced normal, abnormal, with and without X-Factor by using K-Fold cross-validation (5-Folds). This is necessary to avoid over-fitting the model, but it might result in including patients$'$ beats in the training set and reporting on testing set that include the same data which will falsely inflate the measures of accuracy. 5-fold cross validation was performed, since the X-Factor beats (segments) are much less than the normal and abnormal, in which 5-folds is keeping enough X-Factor beats for testing. The second approach, the recordings were split into two balanced training and testing sets, where the recordings in the testing set include almost the same portion of beats/recordings from normal, abnormal, with or without X-factor class. This approach provides a more thorough analysis of the reported classification performance and measures the ability of the trained HMM models to classify any unseen heart sound data.

\subsubsection{Beat-level Classification using 5-Fold Cross-validation (Without X-Factor)}
Table \ref{Table:table9} shows the corresponding results from 5-fold cross validation for a total of 81,498 normal and abnormal beats. We partition the database to include balanced proportions of normal and abnormal beats for both training and testing, note that each fold might not contain the exact amount of recordings compared to the other folds. 
The overall results for the normal/abnormal classification can be seen in the last two rows of the table, showing an average $Se$ of $94.39 \pm 1.22$, $P_+$ of $86.37 \pm 0.9$, $Acc$ of $87\pm0.52$, and $F1$ score of $90.19 \pm 0.26$. The four evaluation metrics ($Se$,$P_+$,$Acc$, and $F1$) . Note that some of normal/abnormal beats are corrupted by varied levels of noise; nevertheless, the total noisy recordings were excluded from this experiment. Moreover, the database does not provide the beat-level cardiologists$'$ labeling. This may result in miss-classification of a beat with noise as abnormal as it can be noticed in $FP$ column (see Table \ref{Table:table9}).

\begin{table*}[!t]
\renewcommand{\arraystretch}{1.3}
\caption{K-Fold (5-Fold) cross validation of Physionet CinC training dataset (Table \ref{Table:table1}) without X-Factor.}
\vspace{0.1 cm}
\label{Table:table9}
\centering
\resizebox{1\textwidth}{!}{
\begin{tabular}{c|cccccccc|cccccccc}

\hline \hline
	\multirow{2}{*}{\begin{tabular}[c]{@{} c@{}}Fold\\iterate \end{tabular}} & \multicolumn{8}{c}{Beat-level without X-Factor class} & \multicolumn{8}{c}{Recording-level without X-Factor class}\\
	\cline{2-17}
	  & $TP$ & $FP$ & $TN$  & $FN$ & $Se$  & $P_+$ & $Acc$ & $F1$ & $TP$ & $FP$ & $TN$  & $FN$ & $Se$  & $P_+$ & $Acc$ & $F1$ \\
	  \hline
	1 & 3079 & 1741 & 11290 & 190  & 94.19 & 86.64 & 88.15 & 90.26 & 516 & 310 & 1899 & 35  & 93.67 & 85.97 & 87.51 & 89.65\\
	2 & 3040 & 1721 & 11309 & 229  & 92.99 & 86.79 & 88.04 & 89.79 & 512 & 362 & 1864 & 35  & 93.60 & 83.74 & 85.68 & 88.40\\
	3 & 3134 & 1887 & 11143 & 135  & 95.87 & 85.52 & 87.59 & 90.40 & 530 & 409 & 1818 & 17  & 96.89 & 81.63 & 84.64 & 88.61\\
	4 & 3118 & 1903 & 11128 & 151  & 95.38 & 85.40 & 87.40 & 90.11 & 538 & 392 & 1827 & 15  & 97.29 & 82.33 & 85.32 & 89.19\\
	5 & 3058 & 1627 & 11403 &  212 & 93.52 & 87.51 & 88.72 & 90.42 & 505 & 325 & 1884 & 40  & 92.66 & 85.29 & 86.75 & 88.82\\
	 \hline
	Mean & 3086 & 1776 & 11255 & 183  & 94.39 & 86.37 & 87.98 & 90.19 & 521 & 360 & 1858 & 28  & 94.82 & 83.79 & 85.98 & 88.93\\
	SD$^\dagger$  & 40 & 117 & 117 & 40  & 1.22 & 0.90 & 0.52 & 0.26 & 13 & 42 & 35 & 12  & 2.11 & 1.85 & 1.14 & 0.50\\
	 \hline \hline
	 \multicolumn{9}{m{3cm}}{Standard deviation$^\dagger$}
\end{tabular}}
\vspace{-0.2in}
\end{table*}

\subsubsection{Beat-level Classification using 5-Fold Cross-validation (With X-Factor)}
The heart beats assigned to X-Factor were used together with the normal and abnormal classes. Three HMM models were trained for normal, abnormal, X-Factor class. The objective of this experiment is to test the ability of the proposed method to automatically reject the beats which labeled as unsure, this is a challenging task in the biomedical signal analysis. The metrics used to evaluate the classification performance are $Se$, $P_+$, $Acc$, and $F1$ score. In order to confirm the overall performance of the beats being classified as normal or abnormal with the existence of X-Factor class, a modified performance measure metrics as defined in equations (\ref{Eqn:eqn16}), (\ref{Eqn:eqn17}), (\ref{Eqn:eqn18}), and (\ref{Eqn:eqn19}) were used. The confusion matrix is obtained for each of 5-fold cross-validation dataset, in which the reference beat labels A-good represent the beats confirmed to be abnormal and A-poor refers to those beats considered as unsure (X-Factor). The incorporation of X-Factor class came at cost of almost 13.3\% of the X-Factor beats goes to abnormal class and 7.6\% classified as normal. Table \ref{Table:table11} shows the average performance of the 5-fold cross validation, the method achieved an average $Se$ of $83.82\pm 2.47$, $81.63\pm 1.4$ $Sp$, $82.73 \pm 1.67$ $Acc$, and $82.7 \pm 1.66$ $F1$ score. The small values of the standard deviations in the last row indicate consistent results across the 5-folds. 

\begin{table}[!t]
\renewcommand{\arraystretch}{1.3}
\caption{K-Fold (5-Fold) cross validation performance for Physionet CinC training dataset (Table \ref{Table:table1}) with X-Factor.}
\vspace{-0.1 cm}
\label{Table:table11}
\centering
\resizebox{0.6\textwidth}{!}{
\begin{tabular}{c|cccc|cccc}
\hline \hline
	\multirow{2}{*}{\begin{tabular}[c]{@{} c@{}}Fold\\iterate \end{tabular}} & \multicolumn{4}{c}{Beat-level with X-Factor} & \multicolumn{4}{c}{Recording-level with X-Factor class}\\
	\cline{2-9}
	  & $Se$  & $Sp$ & $MAcc$ & $F1$ & $Se$  & $Sp$ & $MAcc$ & $F1$ \\
	  \hline
	1 & 81.45 & 82.12 & 81.78 & 76.25 & 77.93 & 81.59 & 79.76 & 76.57\\
	2 & 82.62 & 79.63 & 81.12 & 76.46 & 81.29 & 78.78 & 80.04 & 76.29\\
	3 & 85.47 & 81.07 & 83.27 & 77.15 & 81.88 & 79.22 & 80.55 & 75.56\\
	4 & 87.31 & 83.43 & 85.37 & 77.36 & 86.76 & 80.66 & 83.71 & 77.48\\
	5 & 82.24 & 81.92 & 82.08 & 77.83 & 79.87 & 80.44 & 80.16 & 76.51\\
	 \hline
	Mean & 83.82 & 81.63 & 82.73 & 77.01 & 81.55 & 80.14 & 80.84 & 76.48\\
	SD$^\dagger$  & 2.47 & 1.40 & 1.67 & 0.65 & 3.29 & 1.13 & 1.63 & 0.69\\
	 \hline \hline
	 \multicolumn{5}{m{3cm}}{standard deviation$^\dagger$}
\end{tabular}}
\vspace{-0.2in}
\end{table}

\subsubsection{Recording-level Classification using 5-Fold Cross-validation (Without X-Factor)}
In this experiment, the whole heart sound recording was classified either as normal or abnormal in discarding the inter-beat classification. Table \ref{Table:table9} shows the detailed performance of 5-fold cross validation on the selected balanced normal-abnormal dataset. The FP rate for detecting the abnormal recordings is showing that almost 16.23\% of the normal recordings were classified as abnormal which increases the probability of false classification. However, the proposed method obtains a $Se$ of $94.82\pm 2.11$, $83.79\pm 1.85$ $P_+$, $85.98\pm 1.14$ $Acc$, and $88.98\pm 0.5$ $F1$ score. Compared to beat-level classification performance in Table \ref{Table:table9}, the performance shows a slightly drop for record-level classification. This indicates that some of the recordings may be considered as abnormal based on the existence of abnormality in some beats while other beats are still holding the normal morphologies.

\subsubsection{Recording-level Classification using 5-Fold Cross-validation (With X-Factor)}
In the recording-based classification with X-Factor class, each recording labeled as unsure was considered as X-Factor. Since the X-Factor recordings do not include the fundamental heart sounds ($S1$, Systole, $S2$, Diastole), the X-Factor recordings are segmented using non-overlap window of size one-$second$. This segmentation was considered an equivalent to the complete heart beat cycle ($S1$ sound to end of diastole) in the normal or abnormal recordings. Compared to the beat-level classification without incorporating X-Factor class, we can observe that the average $Se$ dropped from $94.82\pm 2.11$ to $81.55±\pm 3.29$ (see Table \ref{Table:table11}), so as the other metrics. This drop in performance occurs due to the recordings considered as X-Factor may still holds underlying dynamics of the heart sounds in some portions, which in turn miss-classified as normal or abnormal.

\subsubsection{Beat-level Classification using Leave-one-out (unseen) Cross-validation (Without X-Factor)}
Each dataset (DS-\textit{a} to DS-\textit{e}) is split into train and test set (see Table \ref{Table:table2}) where the testing set contains a balanced and totally unseen recordings to the trained classifier. The HMM classification performance was investigated at both the beat-level and recording-level with or without considering the X-Factor class.
The training and testing sets are shown in Table \ref{Table:table2}, where a total of 1438 normal and abnormal recordings were assigned to training dataset and 1434 normal and abnormal recordings were assigned to testing dataset. Table \ref{Table:table15} shows the performance for abnormal beat detection. Our method achieved an overall accuracy of 86.79\% compared to 87.98\% for 5-fold cross-validation. This provides an evidence that the trained HMM models can achieve almost similar accuracies for both seen and unseen heartbeat testing sets.

\begin{table}[!t]
\renewcommand{\arraystretch}{1.3}
\vspace{0.2in}
\caption{Classification performance for unseen testing set (Table \ref{Table:table2}).}
\vspace{0.1 cm}
\label{Table:table15}
\centering
\resizebox{0.7\textwidth}{!}{
\begin{tabular}{ccccc|cccc}
\hline \hline
	\multirow{2}{*}{\begin{tabular}[c]{@{} c@{}}Classification\\approach \end{tabular}} & \multicolumn{4}{c}{Without X-Factor} & \multicolumn{4}{c}{With X-Factor class}\\
	\cline{2-9}
	& $Se$  & $P_+$ & $Acc$ & $F1$ & $Se$  & $Sp$ & $MAcc$ & $F1$\\
	\hline
	Beat-level & 91.45 & 85.97 & 87.07 & 88.63 & 81.50 & 83.97 & 82.74 & 75.60 \\
	Recording-level & 92.25 & 81.74 & 83.82 & 86.68 & 78.92 & 79.65 & 79.28 & 74.47 \\
	\hline \hline
\end{tabular}}
\vspace{-0.1in}
\end{table}

\subsubsection{Beat-level Classification using Leave-one-out (unseen) Cross-validation (With X-Factor)}
Merging the X-Factor train-test dataset in Table \ref{Table:table8} with the normal-abnormal train-test datasets in Table \ref{Table:table2}, a total of 43123/1576 segments/recordings were assigned to training dataset and 43203/1575 segments/recordings were assigned to testing dataset. The modified $Se$, $P_+$, and $Acc$ were calculated as defined in \cite{Springer2016}, while $F1$ was found using equation (\ref{Eqn:eqn19}). Including the X-Factor, the resulting $Se$ was almost similar compared to the $Se$ discarding X-Factor class; however, the $F1$ score dropped by 13.03\% (see Table \ref{Table:table15}). This is mainly due to the significant imbalanced data classes, where X-Factor contains much smaller amount of data compared to normal and abnormal classes.

\subsubsection{Recording-level Classification using Leave-one-out (unseen) Cross-validation (Without X-Factor)}
The HMM models were trained using 1150 normal and 288 abnormal recordings. The HMM performance was evaluated on the totally unseen testing set containing 1150 normal and 284 abnormal. Table \ref{Table:table15} summarizes the confusion matrix and overall classification performance for the heart sound abnormal recordings detection. We can see an improvement in $Se$ by 3.38\% compared to the beat-level classification. However, the $F1$ score dropped by 1.95\%. This is because of a lower $P_+$ as a trade-off in the increment of $Se$.

\subsubsection{Recording-level Classification using Leave-one-out (unseen) Cross-validation (With X-Factor)}
A total 1150, 288, and 138 normal, abnormal, and X-Factor recordings respectively were used to train the HMM models. The HMM performance was evaluated on the totally unseen testing set containing 1150 normal, 284 abnormal, and 141 X-Factor. The classification confusion matrix is obtained to compute the performance of heart sound recordings using unseen testing set incorporating X-Factor class, as shown in Table \ref{Table:table15}. We can see a significant drop in $Sp$ which in turn affects the $F1$ score, the classification of heart sound recordings with the inclusion of X-Factor class shows the lowest $F1$ score while maintaining abnormal class $Se$. 

\section{Conclusion}
\label{sec:prior}

We have developed a Markov-switching linear dynamic model of the piece-wise AR process for heart sound segmentation. Results showed that the fusion of SKF and Viterbi algorithm was able to achieve remarkable segmentation accuracy on a challenging dataset. This work focuses on modeling of raw heart sound signals. Future work will consider an extension of the currently proposed model to a multivariate case for modeling multi-dimensional feature vectors extracted from a raw heart sound as in logistic regression model with multivariate normal (MVN) distribution a state-of-the-art method \cite{Springer2016} for heart-sound segmentation. We also investigated the classification performance
of the MFCC-based continuous density HMM which models---not only the normal and abnormal morphologies of heart sound signals but also morphologies considered as unclassifiable or unknown morphologies (denoted as X-Factor). The HMM classification performance was examined with and without incorporating the X-Factor on the 2016 Physionet/CinC Challenge database. Our proposed method shows the best gross $F1$ score of 90.19 and 82.7 on abnormal beat classification with and without incorporating the X-Factor mode respectively.

\newpage

\bibliographystyle{IEEEbib}
\bibliography{Ref-arxiv-reduced}

\end{document}